İktisat Politikası Araştırmaları Dergisi
**Journal of Economic Policy Researches**

Research Article        🔓 Open Access

# Measuring and Rating Socioeconomic Disparities among Provinces: A Case of Türkiye

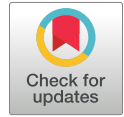

Emre Akusta [1] 🆔 ✉

[1] Kırklareli University, Faculty of Economics and Administrative Sciences, Department of Economics, Kırklareli, Türkiye

**Abstract**

Regional disparities in the economic and social structures of countries have a great impact on their development levels. In geographically, culturally and economically diverse countries like Türkiye, determining the socioeconomic status of the provinces and regional differences is an important step for planning and implementing effective policies. Therefore, this study aims to determine the socioeconomic disparities of the provinces in Türkiye. For this purpose, a socioeconomic development index covering the economic and social dimensions of 81 provinces was constructed. For the index, 16 different indicators representing economic and social factors were used. These indicators were converted into indices using the Min-Max normalization method and Principal Component Analysis. Afterwards, using these indices, the provinces were divided into groups using the K-Means clustering algorithm and the Elbow method. In the last part of the study, the results are presented in a visual format using Scatter Plots, clustering maps and QGIS mapping tools. The results of the study show that 2 of the 81 provinces in Türkiye have very high, 30 high, 25 medium and 24 low socioeconomic indices. Istanbul and Ankara have very high socioeconomic status. In general, the provinces in western Türkiye have a high socioeconomic index, while the provinces in eastern and southeastern Anatolia face serious challenges in terms of socioeconomic indicators.

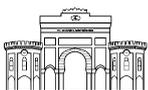

# Measuring and Rating Socioeconomic Disparities among Provinces: A Case of Türkiye

Regional disparities in the economic and social structures of countries have a significant impact on their development levels. Regional development disparities constitute an important problem not only for Türkiye but also for other countries of the world. In Türkiye, as in every country, regional development disparities lead to inequalities in economic and social opportunities and deepen socioeconomic development gaps. Analyzing these disparities plays a vital role in determining and implementing a balanced development strategy (Servi & Erişoğlu, 2020).

In line with the global trend, socioeconomic development levels have been an important subject of study in Türkiye since the 1960s. In the 18th century, an upward trend in socioeconomic development disparities started with the industrial revolution, and these disparities became more pronounced after the Second World War. In the case of Türkiye, the socioeconomic development gaps between geographical regions and provinces have been the focus of various research and studies. In addition, various strategies and projects are being developed at the national and regional levels to overcome this problem, and regional development and investment incentives are offered (Özkubat & Selim, 2019). In Türkiye, under the coordination of the Ministry of Industry and Technology, strategies for regional development and reducing development gaps are implemented in cooperation with the public, private sector and non-governmental organizations.

There are significant disparities among the provinces in Türkiye in terms of their development levels. These disparities have a complex structure shaped by the interaction of various factors. Many variables such as the distribution of economic activities, educational opportunities, infrastructure and superstructure investments, geographical and demographic characteristics are the main factors determining the development disparities between provinces. In order to overcome this problem, regional development gaps must first be determined. While the measurement of development gaps between provinces or regions was initially considered equivalent to per capita income, it is now considered from a multidimensional perspective under the influence of economic, social, demographic, technological, cultural and environmental factors. This multidimensional approach is critical for regional planning and development strategies (Tunç et al., 2009). Therefore, decision-making and classification techniques come to the forefront. Decision-making and classification techniques are important tools for individuals and organizations to achieve various goals and objectives. These techniques provide effective methods for clustering and analyzing issues such as socioeconomic development.

Analyzing the development levels among provinces is an important step for reducing regional inequalities and development. Therefore, this study aims to determine the socioeconomic disparities of the provinces in Türkiye. For this purpose, a socioeconomic development index covering the economic and social dimensions of 81 provinces was developed. This index is calculated for the period 2000-2022 and the provinces are compared over time. Clustering analyses were carried out based on these indices.

This study contributes to the literature in at least five ways: (1) This study develops a new socioeconomic development index that covers 81 provinces of Türkiye and examines the economic and social dimensions both separately and as a whole. In particular, the omission of infrastructure indicators such as sewerage services alongside electricity consumption, which is often neglected in many studies, has been addressed by including these indicators in the index calculations. (2) The developed index is calculated to cover the last two decades of Türkiye. This allows for a dynamic analysis by comparing the performance of the provinces over time. (3) The most recent data set available for the provinces in Türkiye is used. This provides a real-time and up-to-date perspective on the current state of Türkiye's socioeconomic situation. (4) Methodological diversity is provided by using various statistical techniques such as Min-Max normalization method, principal component analysis, and k-means algorithm. (5) Visualization of research findings not only makes the results easier to understand but also sets an example for the use of innovative and effective methods in the presentation of such analyses.

The rest of the paper is organized as follows. Section 2 presents the literature review, Section 3 presents the data and methodology, Section 4 presents the results and discussion, and Section 5 presents the conclusions.

## Literature Review

Clustering provinces in a country based on socioeconomic indicators is important for comprehending economic, social and cultural disparities as well as regional development levels. In this regard, various studies in literature have investigated the development level of provinces by using different statistical methods and multidimensional data analysis techniques. Among the international studies, Cziraky et al. (2002) analyzed the development levels of 545 municipalities in Croatia in 1999 and 2000 using a method based on the covariance structure. Municipal data were collected on economic, structural and demographic dimensions, and confirmatory factor and principal component analyses were applied. As a result of these analyses, the regional development levels were determined. Soares et al. (2003) analyzed the socioeconomic development status of the provinces in Portugal with 33 indicators, and these variables were reduced to 9 factors. By cluster analysis, Portugal was separated into four regions. Buonanno and Montolio (2008) investigated the impact of socioeconomic and demographic factors on crime rates in Spain. Kronthaler (2003) examined the regional development of Eastern and Western Germany and found that some East German regions are economically equal to or better than West German regions. In contrast, Chairat et al. (2015) focused on clustering provinces in Thailand. The result of the study emphasizes various factors influencing clustering, including functional and micro-based concepts. Moreover, studies such as Li et al. (2014) and Han et al. (2020) clustered cities based on factors such as technological elements, geographical factors and human capital. Studies such as Saveleva (2022) and Dumayas (2023) examined the structure of city systems in provinces and their impact on socioeconomic conditions. The results show that the organization of cities can impact socioeconomic conditions. Moreover, studies such as Turgel et al. (2020) and Kansakar and Gupta (2023) address the policy implications of clustering, the development of cluster-based policies, and the legal framework for cluster policies in different countries.

In addition to international studies, comprehending and evaluating the socioeconomic structure of the provinces in Türkiye holds a significant position in academic literature. Researchers have employed diverse statistical methods and socioeconomic indicators to classify and compare the development levels of provinces and regions. Among these studies, Filiz (2005) clustered the provinces in Türkiye into 7 different clusters based on 16 socioeconomic variables using the k-means clustering technique. Albayrak (2005) analyzed the socioeconomic development levels of provinces by factor analysis and identified 8 factors. This study contributes to the multidimensional analysis of socioeconomic structures. Kaygısız et al. (2005) separated the provinces in Türkiye into 5 different clusters by combining Path analysis and cluster analysis. Altıparmak and Özdemir (2005) comparatively analyzed the socioeconomic development levels of provinces using multivariate statistical techniques. Factor analysis based on health, education and economic indicators enabled the evaluation of the socioeconomic structures of provinces. Kılıç et al. (2011) examined the regional similarities and disparities of the provinces in terms of socioeconomic indicators. This study revealed the similarities and disparities of the provinces based on their social and cultural characteristics as well as their economic structures.

Çakmak and Örkçü (2016) analyzed the provinces in Türkiye using socioeconomic indicators such as health, education and economy with data envelopment analysis. The provinces of Istanbul, Kocaeli, Bayburt, Şırnak and Ardahan were determined as effective provinces. Arı and Hüyüktepe (2019) examined the socioeconomic development status of the provinces in Türkiye by using principal component analysis, fuzzy clustering and multidimensional scaling analysis on economic, social and cultural indicators. Özkubat and Selim (2019) constructed development indices for various economic, physical and social indicators of the provinces in Türkiye. It has been reported that the provinces in the Marmara region generally have high development index values. Servi and Erişoğlu (2020) compared the provinces in Türkiye by clustering them based on their development levels using some socioeconomic indicators. As a result of the study, the provinces in Türkiye were clustered into 4 different development levels. The results show that Istanbul, Izmir and Ankara are in the group with the highest level of development. In addition, the province with the highest level of development was determined as Istanbul. Karadaş and Erilli (2023) aimed to cluster the provinces in Türkiye in terms of socioeconomic indicators using gray cluster analysis. Because of the analysis, Türkiye's socioeconomic development map was drawn. In addition, the Ministry of Industry and Technology prepares "Socioeconomic Development Ranking of Provinces and Regions" (SEGE) reports on the development level of provinces. However, the most recent version of these province-based reports is from 2017. This means that the SEGE report series ended in 2017.

In addition to these studies, the provinces in Türkiye have also been clustered based on various indicators. For instance, Çelik (2013) and Tekin (2015) clustered provinces in terms of health indicators. Çelik (2015) analyzed the impact of agriculture and livestock on the regional economy by analyzing livestock data. Alpaykut (2017) investigated the quality of life and well-being by provinces in Türkiye. The results of the study show that metropolitan cities such as Istanbul, Ankara and Izmir are the provinces that offer the best life. Uysal et al. (2017) analyzed the disparities and similarities of the provinces in Türkiye based on their life index values. Doğrul and Çelikkol (2017) determined in which provinces the creative class in Türkiye is concentrated and which province groups show similar characteristics by cluster analysis. Kandemir (2018) separated provinces into two clusters in terms of tourism using the fuzzy clustering method. Finally, Birkalan and Bay (2022) analyzed Türkiye's regions with Ward clustering method under the assumption of a regional minimum wage. This study addresses the potential impacts of economic policies on regional development.

Generally, studies in the literature reveal that there is a diversity of methodologies and variables used in analyzing the socioeconomic development levels of the provinces in Türkiye. Moreover, it reveals the importance of addressing the socioeconomic development levels of provinces in a multidimensional perspective, including economic, social, cultural and environmental factors. However, studies in literature generally conduct a static analysis based on the data of a specific year. With the index developed in this study, the socioeconomic status of the provinces in Türkiye is analyzed over the last 20 years. This enables a more dynamic analysis by comparing the socioeconomic status of the provinces over time. Moreover, the existing literature generally focuses on demographic variables, quantitative indicators related to education, access to health services, and economic indicators. This study, in addition to these variables, also includes access to infrastructure services and social structure variables. This allows for a multidimensional analysis of the socioeconomic development levels of the provinces in Türkiye.

## Data and Methodology

### Model Specifications and Data

This study aims to determine the socioeconomic development disparities of the provinces in Türkiye. For this purpose, a socioeconomic development index is developed for Türkiye and 81 provinces of Türkiye are analyzed. Moreover, the developed index is calculated for the last 20 years of Türkiye and allows for a dynamic analysis by comparing the situation of provinces over periods. The Turkish Statistical Institute (TurkStat) data were used in the study. In order to work with the most up-to-date data and to determine a consensus period in the data set, data between 2000 and 2022 were used. While Min-Max normalization method and the Principal Component Analysis (PCA) method were used to construct the indices, the k-means algorithm was used for the clustering analysis. Compared to other index construction methods, PCA effectively assimilates the main sources of variance in the data sets, revealing the most informative features. This method is particularly useful for high-dimensional and high-variable data sets and reduces the model complexity. PCA improves the efficiency of k-means cluster analysis. This is because the k-means algorithm performs better on data sets with homogeneous variance and reduced dimensions with PCA. This method achieves homogeneous and well-defined clusters by iteratively assigning data points to groups with similar characteristics. The combination of these two methods both facilitate the interpretation of the data and enables more robust analyses. This helps to achieve clear and focused results that address the research questions. Following the index calculation step, cluster analyses were conducted separately on the basis of economic, social and socioeconomic indicators. The Elbow Method was used to determine the optimum number of clusters. The elbow method determines the optimum number of clusters by identifying the point where the rate of decrease in the total variance between clusters is the highest. As a result of this analysis, the provinces were assigned to four different clusters for each indicator. Scatter plots, cluster maps, and QGIS mapping tools were used for visualization. The visualization phase increases the comprehensibility of the research findings.

The indicators used in our study were compiled from the literature (e.g. Filiz, 2005; Kılıç et al., 2011; Çakmak & Örkçü, 2016; Özkubat & Selim, 2019; Servi & Erişoğlu, 2020; Karadaş & Erilli, 2023). Moreover, the choice of variables used aims to comprehensively analyze the socioeconomic disparities of provinces and emphasize dimensions that are less discussed in the literature. These variables include economic performance (GDP, number of vehicles, crop and animal production values, exports), social structure (housing sales, net migration rate, age dependency ratio, population growth rate), access to health services (number of hospital beds, number of doctors, infant mortality rate), education level (primary school enrollment rate, percentage of university graduates), and infrastructure (electricity consumption, sewerage service).

While previous studies have focused more on demographic variables, numerical indicators related to education, access to health services and general economic indicators, our study includes access to infrastructure services and social structure variables. In particular, infrastructure indicators such as electricity consumption and sewerage services are important and often neglected aspects of quality of life and sustainable development. In addition, agricultural and livestock values better reflect the economic diversity and structural characteristics of local economies. Integrating these variables allows us to assess the socioeconomic profiles of the provinces in a more detailed and multidimensional way. This approach is one of the main elements that differentiates our study from existing literature and makes it unique. The descriptive statistics of the dataset are given as follows.

**Table 1**
*Descriptive Statistics*

| | Variables | Description | Mean | Mdn. | S. D. | Min. | Max. |
|---|---|---|---|---|---|---|---|
| Economics | GDP | Per capita ($) | 6849.1 | 6438.5 | 2919.3 | 1950.0 | 18269.0 |
| | Number of cars | Per thousand people | 101.1 | 96.5 | 58.6 | 8.0 | 301.0 |
| | Value of plant production | Total (thousand $) | 390168.0 | 239488.3 | 512215.5 | 2975.3 | 5145056.0 |
| | Value of animal production | Total (thousand $) | 137030.2 | 76166.8 | 193400.0 | 1892.5 | 1648956.0 |
| | Housing sales | Total | 16924.6 | 7102.5 | 32991.8 | 130.0 | 265098.0 |
| | Electricity consumption | Total per capita (kWh) | 2550.2 | 2138.0 | 1707.3 | 443.0 | 11004.0 |
| | Net Exports | Total (thousand $) | −562713.0 | 25282.0 | 5560981.6 | −64479867.0 | 3160138.0 |
| | Net migration rate | Per mille (‰) | −3.3 | −2.1 | 45849 | −81.0 | 52.6 |
| | Age dependency ratio | Total percentage (%) | 52.1 | 49.5 | 45698 | 37.4 | 93.7 |
| | Population growth rate | Annual per mille (‰) | 45782 | 45874 | 45676 | −149.3 | 109.2 |
| Social | Number of hospital beds | Per hundred thousand people | 268.6 | 265.5 | 95.9 | 40.0 | 607.0 |
| | Number of physicians | Per thousand people | 45778 | 1.0 | 0.6 | 1.0 | 4.0 |
| | Infant mortality rate | Per mille (‰) | 45879 | 45726 | 45811 | 45719 | 45859 |
| | Primary school enrollment rate | Net percentage (%) | 93.6 | 93.5 | 4.0 | 59.0 | 99.9 |
| | Sewerage service | Municipal population (%) | 85.5 | 91.0 | 45730 | 23.0 | 100.0 |
| | Number of colleges graduates | Percentage (%) | 45699 | 45879 | 45782 | 45690 | 45742 |

*Note:* S.D., Min, Max, and Mdn denote the standard deviation, minimum, maximum, and median, respectively.

Table 1 shows the descriptive statistics of the dataset. Descriptive statistics summarize the characteristics of the variables in the dataset, such as central tendency, dispersion, and variance. The data reveal that the GDP per capita has a mean of 6849.1 dollars and a standard deviation of 2919.3 dollars. The number of automobiles is 101.1 per thousand people. The high standard deviation indicates that automobile ownership is widespread, but there are significant differences in the distribution among provinces. Crop and animal production values are high; however, the standard deviation of crop production is higher than that of animal production. This indicates that there is more variability in crop production. However, housing sales and per capita electricity consumption fluctuate within a wide range. This reveals that there are significant differences between the provinces in housing sales and electricity consumption. Negative values in exports and the net migration rate are also noteworthy. The age dependency ratio is quite stable, and the high variability in the population growth rate reflects the instability of the population across provinces. The number of hospital beds and the number of doctors are at average values, while the schooling rate is high. This situation shows that significant achievements have been made in the field of education. Similarly, sewerage services and faculty graduation rates are generally high, indicating that a certain level of basic infrastructure and higher education has been achieved.

## Index Techniques

Multi-criteria decision-making methods have an important position in analyzing several variables belonging to objects. In this process, the number of variables and the relationships between variables affect the quality and clarity of the analysis. Principal Component Analysis (PCA), a multi-criteria decision-making method, is a well-established and widely used method that aims to explain a large number of variables with fewer independent variables (Jolliffe, 2002). In other words, PCA makes it possible to obtain more simple and clear results by reducing the number of variables in data sets. PCA was first developed by Pearson (1901) and later improved by Hotelling (1933). PCA, which has found application in many disciplines over time, simplifies the relationships between variables, making analysis processes more understandable and manageable.

PCA obtains fewer variables from several variables that can be considered a combination of variables. This process ensures that no loss of statistical data and that subjective weighting is avoided. As a result, p variables are transformed into k (k ≤ p) new linear and independent variables (Jolliffe, 2002:167). However, large differences in the value ranges of different variables can lead to attributes with smaller values being ignored in the analysis process. Therefore, the first step to be taken when starting the principal component analysis is to normalize the variables to be subjected to the analysis. The normalization process prevents such imbalances and ensures that all variables contribute equally to the analysis. Data normalization methods include Min-Max, Z-Score, Decimal Scaling, Sigmoidal and Softmax (OECD, 2008). In this study, Min-Max normalization method was preferred for data normalization.

## Cluster Analysis

Cluster analysis is an important statistical method used to group units with similar characteristics. The main objective of this method is to group units in a data set into subgroups that are as similar as possible (Aldenderfer & Blashfield, 1984). The clustering method emerged in the 1930s and gained popularity with the development of computer technologies, especially since the 1960s (Blashfield, 1976). Clustering methods are generally analyzed under two main headings: hierarchical and non-hierarchical. The most widely used non-hierarchical clustering method is the K-Means method (Xu & Wunsch, 2009). The history of cluster analysis dates back to 1939 when R. C. Tryon used it as a systemic analysis method (Tryon, 1939). Similar to multidimensional scaling analysis, cluster analysis is based on mathematical properties that do not require statistical inference (Tatlıdil, 1992). This method has been applied in various disciplines such as biology, sociology, psychiatry and statistics (Everitt et al. 2011).

The k-means clustering method was proposed by MacQueen (1967) and is an algorithm that maximises the similarity between clusters. This method separates the data in the dataset into a set of k groups. The performance of the k-means algorithm depends on factors such as the choice of initial cluster centers and the value of k (Meila & Heckermen, 1998; Mohamed & Çelik, 2022). The steps of the k-means clustering method are summarized below (Steinbach et al., 2000; Karypis et al., 2000; Risheh et al., 2022):

- **Step 1:** Determine the number of clusters (k) to be grouped
- **Step 2:** Determine the initial cluster centers as many as the number of clusters to be grouped
- **Step 3:** Assign the data in the dataset to the cluster whose cluster center is closest to itself
- **Step 4:** Once the clusters are identified (Step 3 is completed), calculate the cluster means and assign the obtained means as new starting centers
- **Step 5:** Repeat Step 3 and Step 4 until the data in the clusters are fixed.

In conclusion, cluster analysis is a widely used method in data science and statistics for the explanation and classification of data. The k-means method is preferred due to its easy applicability and comprehensibility in practice. However, the success of the algorithm depends on critical parameters such as the correct determination of the k value and the selection of the starting points. Incorrect determination of this value may negatively affect the reliability of the analysis results and may lead to incorrect results. In the existing literature, various methods are used to determine this critical value. These methods include the Elbow method, Calinski and Harabasz's (1974) index, Davies and Bouldin's (1979) index, Krzanowski and Lai's (1988) index and Silhouette's (1987) index. In this study, the elbow method is preferred, and the details of this method are given in the rest of the study.

## Elbow Method

The elbow method is a technique used in k-means cluster analysis to determine the optimal value of k, the number of clusters. This method is based on the sum of squares of the distances to the cluster centers, the Sum of Squares within Cluster (WCSS). Within the Elbow method, the WCSS generally decreases as k increases. Nevertheless, the magnitude of this decrease diminishes considerably at a specific point, referred to as the elbow point. The elbow point is the point at which the next cluster does not significantly reduce the total error squares. This point represents the optimal number of clusters (Ketchen & Shook, 1996; Amasyalı & Ersoy, 2008).

The Elbow method provides a visual assessment and is often used to easily observe the change in the rate of decrease of WCSS on a graph. The following are the steps of the Elbow method (Bholowalia & Kumar, 2014; Syakur et al., 2018; Taşçı & Onan, 2016; Coşkun et al., 2021):

- **Step 1:** Calculate the WCSS value for each k starting with k = 1
- **Step 2:** Recalculating the WCSS value by increasing k
- **Step 3:** Determine the point at which the WCSS reduction is significantly reduced (this point at which the reduction is significantly reduced is the elbow point at which the optimal number k is determined).
- **Step 4:** Determine the value of k at which the WCSS is reduced less, and the modelling is not improved by adding an additional cluster.

This method is important to minimize the subjectivity in determining the correct k value and to better comprehend the structure of the clustered data. The Elbow method improves the efficiency of the k-means algorithm by analyzing the WCSS values for different k values depending on the structural characteristics of the dataset. Therefore, it is important to use methods such as Elbow before applying the k-means algorithm.

## Results and Discussion

In the first stage of the analysis, the correlation between the indicators used in the study was examined. The Correlation Matrix Heatmap, shown in Figure 1, visualizes the degree of correlation between different metrics using a colour scale. Correlation values range from −1 to +1: +1 indicates a perfect positive correlation, 0 indicates no correlation and −1 indicates a perfect negative correlation. The colour scale ranges from red to blue, with red shades representing a high positive correlation and blue shades representing a high negative correlation. In this regard, the Correlation Matrix Heatmap provides a quick assessment of the relationships between metrics.

**Figure 1**

*Correlation Matrix Heatmap*

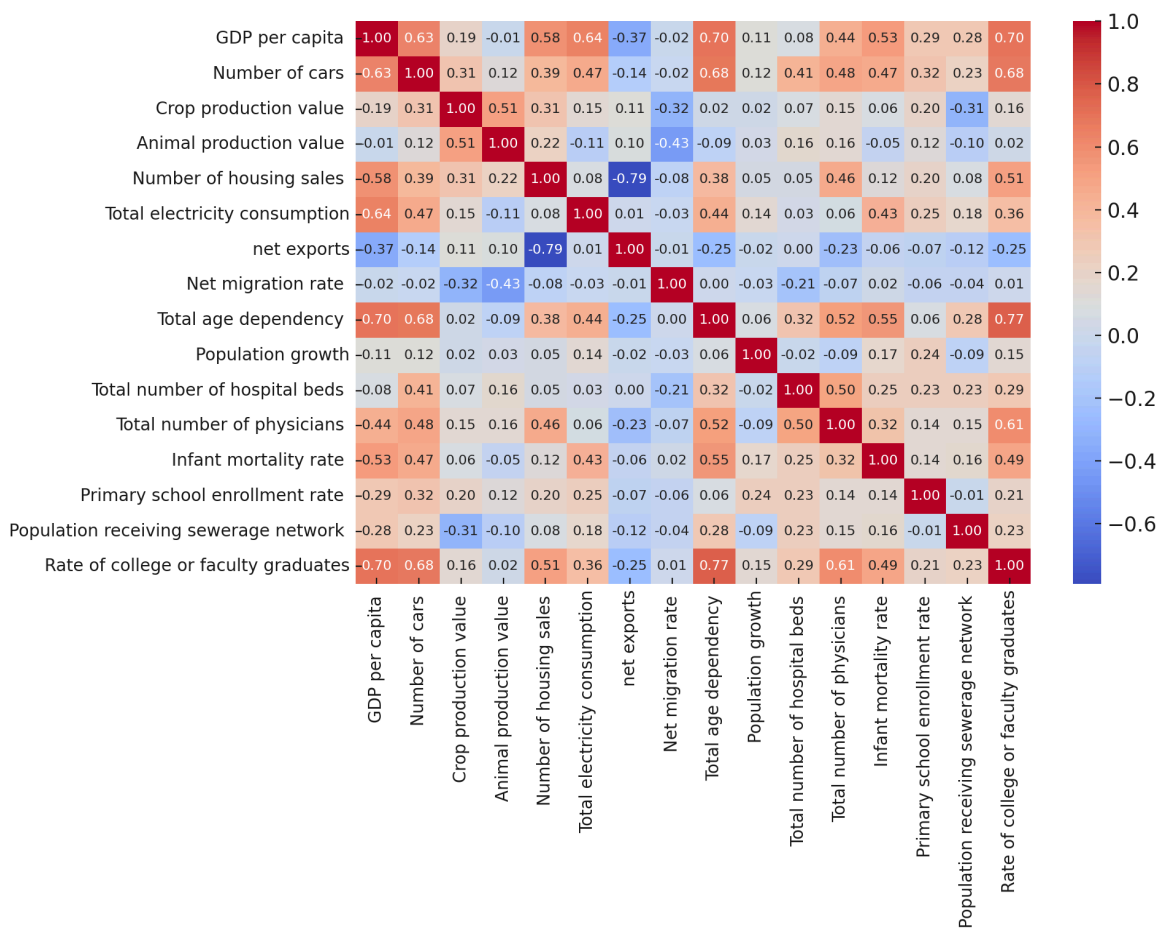

The correlation matrix heat map shown in Figure 1 provides a visual representation of the relationships between the variables used in the study. The analysis reveals that there are significant correlations between some indicators. For instance, the high positive correlation between GDP per capita and the car ownership rate shows that individuals can acquire more vehicles as economic prosperity increases. This means that economic growth has a significant impact on consumption patterns and that car ownership may become a status symbol as the level of welfare increases. Meanwhile, the positive correlation observed between agriculture and livestock production values demonstrates that these two sectors are closely interrelated and generally develop together. This points to an ecosystem where agriculture provides the necessary inputs for livestock production and livestock production adds value to agricultural products. Such a relationship may have important implications for rural development policies and sustainable agricultural practices. There is a positive correlation between education and access to health services. It is known that communities with higher levels of education are more advanced in terms of access to health services and health awareness. This demonstrates that education increases knowledge and skills that can improve individuals' access to health services and their health-related decisions. Therefore, it can be concluded that investments in the education and health sectors have mutually supportive and reinforcing effects. However, the low correlation between net exports and the population growth rate shows that there is no direct relationship between these two indicators. This may indicate that the provinces' foreign trade balance is shaped by factors such as economic policies, global market conditions and production capacity, independent of the population growth rate.

**Figure 2**
*Pair Plot Analysis Results*

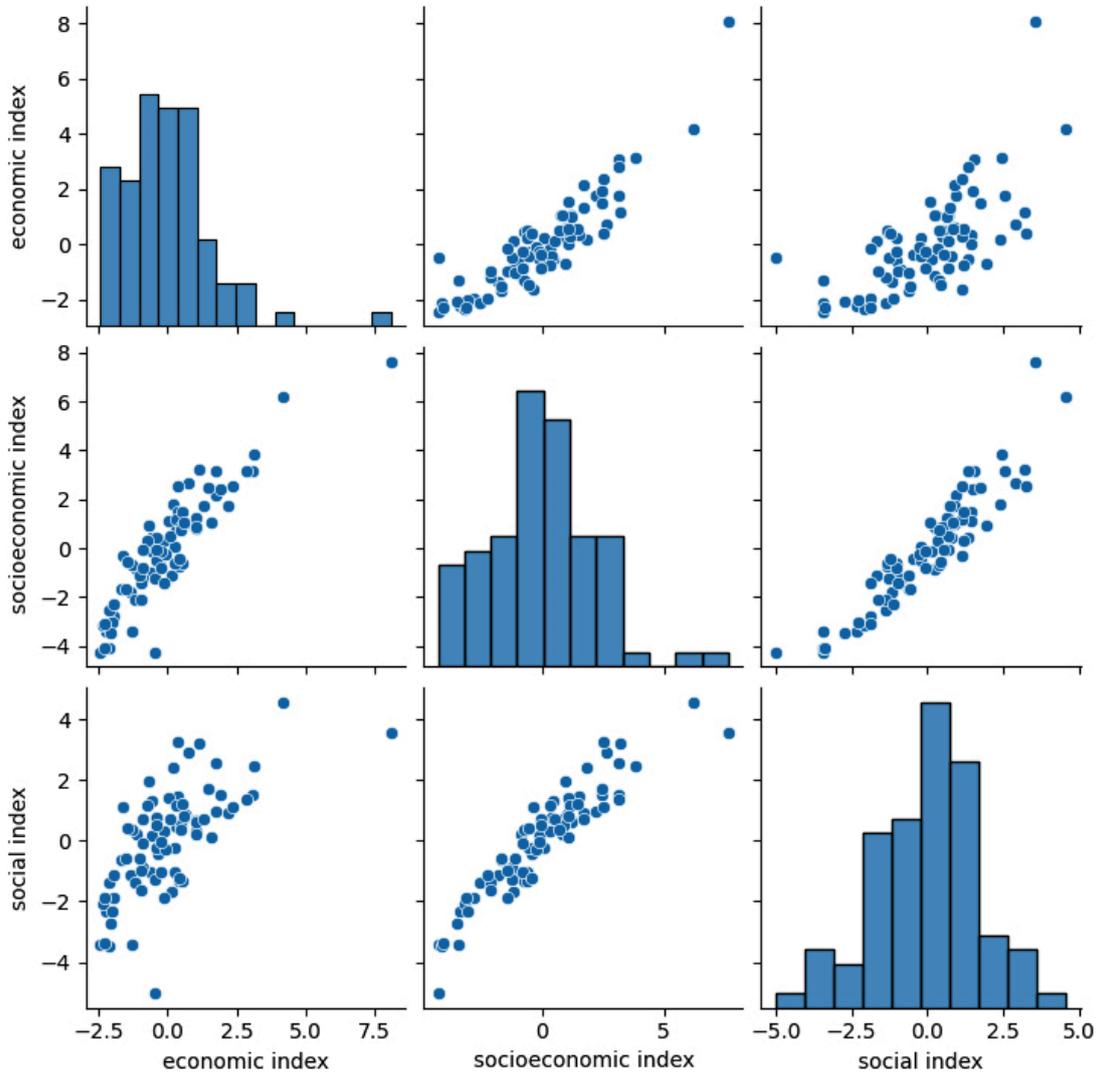

Figure 2 shows the results of the pair plot analysis. The pair plot demonstrates the relationships between several selected metrics in the form of bivariate distributions and correlations. This is a useful method for exploring the potential relationships and trends between variables. Diagonal histograms show the distribution of the economic, social and socioeconomic indices. Each histogram shows the frequency of the data points within a range of values for that index. These histograms are skewed to the right. This indicates that most data are clustered around low values and there are fewer high values. Off-diagonal histograms represent bivariate relationships between different indices.

Figure 2 reveals that the Economic and Socioeconomic Index shows a positive correlation. In other words, as the values of one index increase, the values of the other index also tend to increase. Similarly, the Economic Index and Social Index distribution also indicates a positive correlation. Higher economic index values are associated with higher social index values. Finally, the distribution of the Socioeconomic Index and Social Index also shows a positive correlation. This indicates that cities with a higher socioeconomic index tend to have a higher social index. Overall, the indices are found to be interrelated and tend to move together.

**Figure 3**
*Elbow Method for Determining the Optimal Number of Cluster*

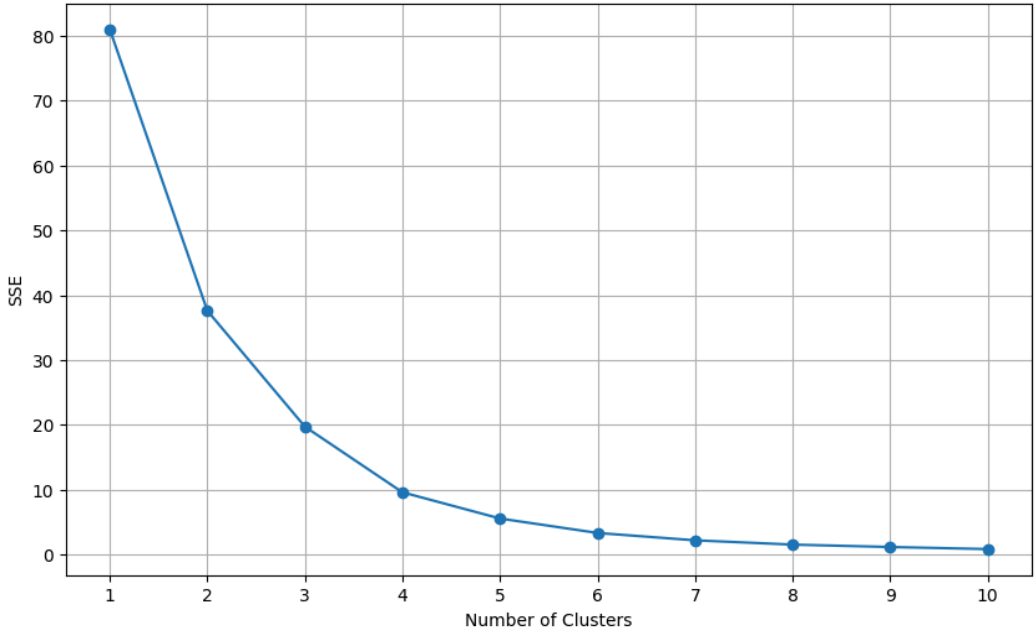

The elbow method is widely used in algorithms such as k-means cluster analysis to determine the optimal number of clusters. However, one of the major challenges of the method is that the elbow point is not always clearly defined. As Kodinariya and Makwana (2013: 92) state, sometimes a clear elbow point may not be observed or multiple elbow points may occur. This can make it difficult to select the optimal number of clusters. In such a case, a decision should be made based on the characteristics of the dataset and the needs of the research question.

Figure 3 shows that the rate of decrease of the Sum of Squared Errors (SSE) values decreases in clusters 3 and 4. This means that these two points can be considered as potential elbow points and the dataset can be divided into 3 or 4 clusters. However, in this study, when the dataset was divided into 3 clusters, provinces with significant differences were in the same cluster. This reduces the sensitivity of the analysis. For example, Istanbul differs significantly from other provinces based on its economic index values. Since Istanbul has a high economic index, it is in the same group as other provinces when evaluated in 3 clusters. However, when the dataset is separated into 4 clusters, Istanbul is alone in the highest index cluster. Therefore, the differences between the provinces become clearer. For these reasons, it was preferred to separate the dataset used in the study into 4 clusters. In this way, the characteristics of each province are represented more clearly and the differences between provinces are shown more distinctly.

**Figure 4**

*Dendrogram of the Indices*

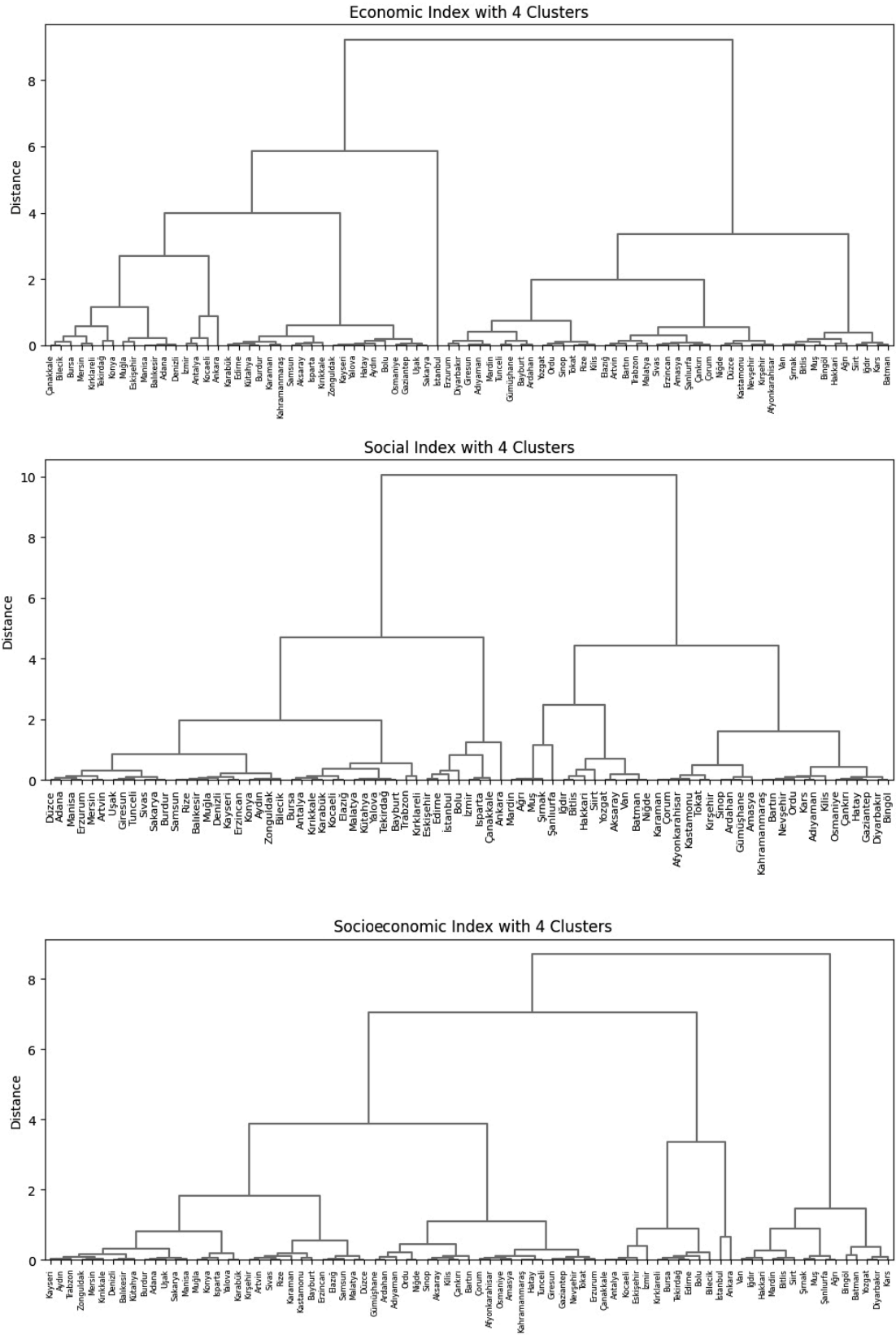

In the dendrogram, the height of the joins (vertical lines) represents the joins between clusters. The height of each union indicates the differences between clusters; longer lines represent greater differences. When two clusters are joined by a high line, this means that they are less similar than clusters joined by shorter lines. The dendrograms in Figure 4 cluster cities by their economic, social and socioeconomic similarities. The distance where the lines meet indicates the point where the clusters meet. A horizontal cut line along the dendrograms can be considered to divide the dendrogram into 4 clusters. The position of this line will depend on the desired number of clusters or a certain threshold distance. Cities connected below this cut-off line are counted in the same cluster. Moving the cut line lower increases the number of clusters, while moving it higher decreases the number of clusters. This results in wider and narrower clusters. This categorization indicates that the cluster with the provinces with the highest index value represents the most economically developed provinces. The provinces in this cluster have the highest GDP, number of vehicles, and number of doctors. In addition, the provinces in this cluster have the lowest infant mortality rate and the highest primary school enrollment rate. This indicates a high overall quality of life. The other clusters are medium and low in terms of socioeconomic indicators.

**Figure 5**

*Economic Indices and Clustering Results of the Provinces*

Figure 5 shows the economic indices and clustering results of the provinces in 2022. Large metropolitan provinces such as Istanbul, Ankara, and Izmir are frequently regarded as the primary drivers of Türkiye's economy. These provinces exhibit high economic index values, with Istanbul standing out as the country's economic epicenter. Istanbul boasts a robust industrial sector, elevated levels of income, and intensive trade operations, contributing to its significantly superior economic index compared to other provinces. Furthermore, Istanbul's per capita electricity consumption surpasses that of other provinces by a considerable margin. This is an indication of the city's dense population and commercial activities. Ankara, another important economic centre, also has a high economic index. The fact that Ankara is the capital and has large-scale agricultural and industrial activities is among the factors contributing to its economic strength.

However, Ankara's lower per capita sales of cars and houses compared to Istanbul is evidence of its lower population density.

Türkiye's provinces in the Eastern Anatolia and Southeastern Anatolia regions generally have lower economic index values. These provinces face more challenges in terms of economic development. Unemployment rates are generally higher and income levels are lower, especially in the regions where agriculture and livestock are widespread. The economic potential in these regions can be increased through strategic investments in agriculture, tourism and alternative energy resources. Provinces in Central Anatolia and the Aegean regions generally have medium economic index values. While these regions operate in various sectors such as agriculture, industry and tourism, they have a generally balanced economic structure. For example, Konya is an important agricultural center, while İzmir attracts attention with its tourism and exports. Although the provinces in the Black Sea region are generally active in sectors such as agriculture and forestry, their economic index values are at a medium level. The economic potential of these regions can be developed with appropriate investments. The results of the analysis clearly indicate that regional disparities should be considered in determining Türkiye's economic development policies. Specific policies and investments should be made for provinces in different regions to reduce regional disparities and ensure economic development. It is also important to regularly update and analyze economic indices to monitor Türkiye's economic performance and determine future strategies.

**Figure 6**

*Social Indices and Clustering Results of the Provinces*

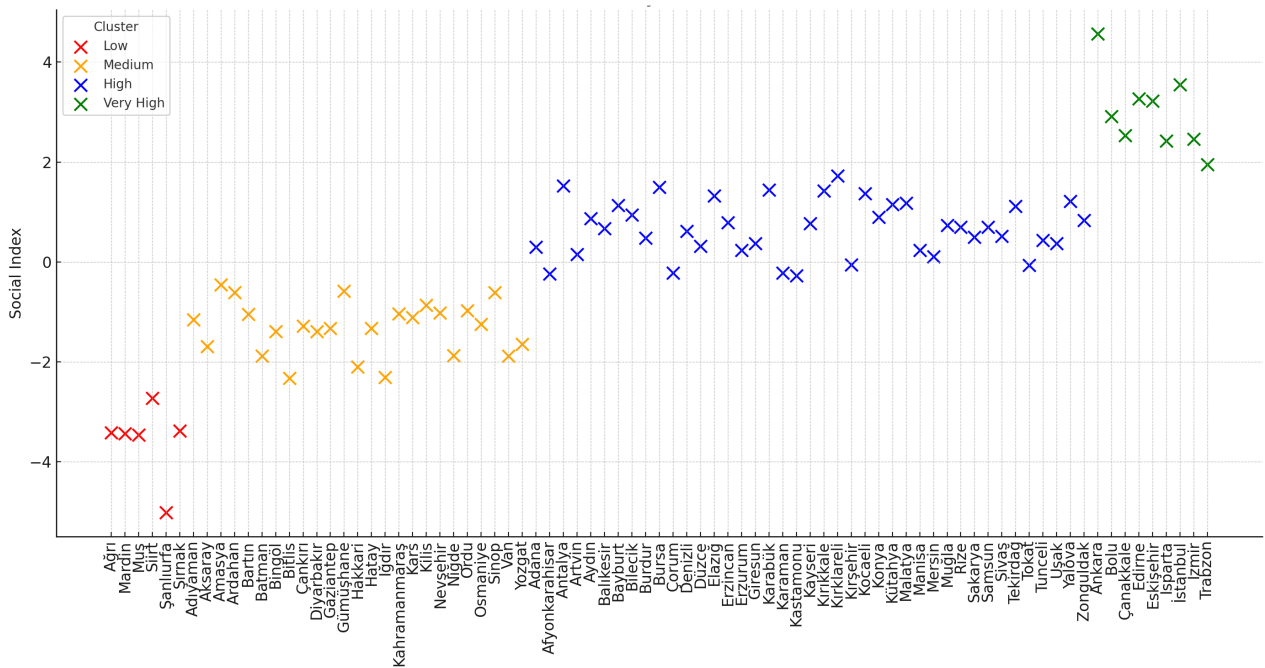

Figure 6 shows the social indices and clustering results of the provinces in 2022. Clustering analysis based on the social index values of provinces revealed that some provinces were more advantageous or disadvantageous in socioeconomic terms compared to others. In other words, the social development levels of the provinces in Türkiye have a heterogeneous structure. In particular, the provinces with the highest social index values are more socially developed and have higher living standards. These provinces stand out with better opportunities in areas such as education, health, and employment. However, provinces with low social index values represent regions that are less socially developed and face various challenges. People living in these provinces are less advantaged in terms of access to education and health services, employment opportunities, and overall quality of life.

Provinces with very high social index values, such as Ankara, Istanbul and Edirne, perform well on indicators such as the number of hospital beds, the number of doctors and primary school enrolment rates. They also have low infant mortality rates and high educational graduation rates. This indicates a relatively higher quality of life and socioeconomic development in these provinces. In particular, Ankara, the capital of Türkiye, has the highest social index value. In addition, the proportion of higher education graduates is also quite high. This shows that Ankara has a high level of education. Provinces with high social index values such as Kırklareli, Konya, Antalya and Denizli are higher than other groups. The social indicator performance of this group is also above average. Meanwhile, provinces with medium and low social index values include provinces with weaker social indicator performance. There are serious problems, especially in areas such as access to health services, education opportunities and social services. This situation emphasizes the need for specific policies and investments to improve the quality of life and social welfare in these regions.

**Figure 7**

*Socioeconomic Indices and Clustering Results of the Provinces*

The clustering results of the socioeconomic indices calculated using the economic and social indices used in the study for 2022 are shown in Figure 7. The results reveal that only 2 of 81 provinces in Türkiye (2.5% of the provinces) are in the very high socioeconomic index cluster. This means that the number of provinces with very high socioeconomic status is very limited across the country. However, 30 provinces (37% of provinces) are in the high socioeconomic index cluster. This indicates that a large proportion of the provinces in Türkiye have a high socioeconomic status, but the number of provinces at the highest level is limited. There are 25 provinces (30.9% of provinces) in the medium socioeconomic index cluster and 24 provinces (29.6% of provinces) in the low socioeconomic index cluster. This distribution shows that 60% of the provinces in Türkiye have medium and low socioeconomic levels and reveals the existence of socioeconomic disparities and imbalances in the country. In the following section of the study, the situation of the provinces by years is analyzed.

**Figure 8**
*All Indices of Provinces*

| Provinces | Economic Index | | | | | | Social Index | | | | | | Socioeconomic Index | | | | | |
|---|---|---|---|---|---|---|---|---|---|---|---|---|---|---|---|---|---|---|
| | 2000 | | 2010 | | 2020 | | 2000 | | 2010 | | 2020 | | 2000 | | 2010 | | 2020 | |
| | Index Value | Cluster | Index Value | Cluster | Index Value | Cluster | Index Value | Cluster | Index Value | Cluster | Index Value | Cluster | Index Value | Cluster | Index Value | Cluster | Index Value | Cluster |
| Adana | 0.85 | M | 0.94 | M | 0.86 | M | 0.44 | M | 0.50 | M | 0.43 | M | 0.50 | M | 0.77 | M | 0.54 | M |
| Adıyaman | -1.46 | L | -1.55 | L | -1.38 | L | -1.63 | L | -1.34 | L | -1.01 | M | -1.18 | M | -1.34 | L | -1.00 | M |
| Afyonkarahisar | -0.08 | M | -0.11 | M | -0.28 | M | -0.25 | M | -0.23 | M | -0.18 | M | -0.13 | M | -0.23 | M | -0.19 | M |
| Ağrı | -2.14 | L | -2.28 | L | -2.48 | L | -3.65 | L | -2.16 | L | -3.63 | L | -2.23 | L | -2.16 | L | -2.52 | L |
| Aksaray | -0.71 | M | -0.46 | M | 0.15 | M | -1.58 | L | -0.78 | M | -1.75 | L | -0.88 | M | -0.78 | M | -0.64 | M |
| Amasya | -0.35 | M | -0.31 | M | -0.24 | M | 0.87 | M | 0.21 | M | -0.49 | M | 0.20 | M | 0.21 | M | -0.30 | M |
| Ankara | 5.35 | VH | 5.15 | VH | 4.84 | VH | 5.88 | VH | 4.26 | VH | 4.00 | VH | 4.31 | VH | 4.26 | VH | 3.67 | VH |
| Antalya | 3.43 | VH | 4.07 | VH | 2.29 | H | 1.40 | H | 2.18 | H | 1.45 | H | 1.85 | H | 2.18 | H | 1.56 | H |
| Ardahan | -2.10 | L | -1.71 | L | -1.60 | L | -1.50 | L | -1.47 | L | -1.35 | L | -1.38 | L | -1.47 | L | -1.22 | M |
| Artvin | -0.88 | M | -0.62 | M | -0.34 | M | 0.55 | M | -0.17 | M | 0.98 | M | -0.12 | M | -0.17 | M | 0.24 | M |
| Aydın | 1.15 | H | 0.70 | M | 0.42 | M | 0.28 | M | 0.73 | M | 0.91 | M | 0.54 | M | 0.73 | M | 0.55 | M |
| Balıkesir | 2.47 | H | 1.93 | H | 0.99 | M | 0.94 | M | 1.04 | H | 0.81 | M | 1.31 | H | 1.04 | H | 0.75 | M |
| Bartın | -0.94 | M | -0.77 | M | -0.55 | M | -0.29 | M | -0.30 | M | 0.67 | M | -0.25 | M | -0.30 | M | 0.03 | M |
| Batman | -1.95 | L | -1.98 | L | -2.13 | L | -2.15 | L | -1.54 | L | -1.89 | L | -1.58 | L | -1.54 | L | -1.67 | L |
| Bayburt | -1.75 | L | -1.56 | L | -1.51 | L | -0.62 | M | -0.69 | M | 0.34 | M | -0.91 | M | -0.69 | M | -0.51 | M |
| Bilecik | 1.03 | H | 1.07 | H | 1.71 | H | 0.98 | M | 0.90 | M | 0.25 | M | 0.77 | M | 0.90 | M | 0.83 | M |
| Bingöl | -2.28 | L | -2.29 | L | -2.07 | L | -1.20 | M | -1.52 | L | -0.98 | M | -1.34 | L | -1.52 | L | -1.28 | M |
| Bitlis | -2.25 | L | -2.36 | L | -2.25 | L | -2.92 | L | -1.89 | L | -2.38 | L | -1.99 | L | -1.89 | L | -1.92 | L |
| Bolu | 1.72 | H | 1.84 | H | 0.81 | M | 1.49 | H | 1.66 | H | 2.45 | H | 1.23 | H | 1.66 | H | 1.33 | H |
| Burdur | 0.41 | M | 0.94 | M | 0.46 | M | 0.72 | M | 0.34 | M | 0.27 | M | 0.43 | M | 0.34 | M | 0.30 | M |
| Bursa | 2.80 | VH | 2.08 | H | 1.83 | H | 0.96 | M | 1.25 | H | 1.57 | H | 1.44 | H | 1.25 | H | 1.41 | H |
| Çanakkale | 1.58 | H | 2.19 | H | 1.91 | H | 0.94 | M | 1.26 | H | 1.70 | H | 0.97 | M | 1.26 | H | 1.50 | H |
| Çankırı | -1.11 | M | -0.96 | M | -0.46 | M | 0.90 | M | -0.48 | M | -1.65 | L | -0.08 | M | -0.48 | M | -0.86 | M |
| Çorum | -0.33 | M | -0.23 | M | -0.38 | M | 0.37 | M | -0.16 | M | 0.79 | M | 0.02 | M | -0.16 | M | 0.15 | M |
| Denizli | 1.24 | H | 0.99 | M | 1.01 | H | 0.16 | M | 0.85 | M | 0.89 | M | 0.53 | M | 0.85 | M | 0.79 | M |
| Diyarbakır | -1.10 | M | -1.38 | L | -1.27 | M | -1.29 | M | -1.11 | M | -1.36 | L | -0.92 | M | -1.11 | M | -1.09 | M |
| Düzce | -0.29 | M | -0.14 | M | -0.17 | M | 0.02 | M | 0.15 | M | 0.04 | M | -0.10 | M | 0.15 | M | -0.06 | M |
| Edirne | 0.35 | M | 0.63 | M | 0.34 | M | 3.10 | VH | 1.68 | H | 2.88 | VH | 1.33 | H | 1.68 | H | 1.30 | H |
| Elazığ | -0.75 | M | -0.76 | M | -0.44 | M | 1.56 | H | 0.33 | M | 1.09 | H | 0.32 | M | 0.33 | M | 0.25 | M |
| Erzincan | -0.70 | M | -0.51 | M | -0.07 | M | 0.57 | M | -0.14 | M | 0.68 | M | -0.05 | M | -0.14 | M | 0.24 | M |
| Erzurum | -1.05 | M | -0.92 | M | -1.02 | M | 0.45 | M | 0.14 | M | -0.48 | M | -0.23 | M | 0.14 | M | -0.63 | M |
| Eskişehir | 1.30 | H | 1.36 | H | 1.26 | H | 3.28 | VH | 1.95 | H | 3.25 | VH | 1.77 | H | 1.95 | H | 1.84 | H |
| Gaziantep | 0.02 | M | -0.36 | M | 0.35 | M | -1.42 | L | -0.76 | M | -1.67 | L | -0.39 | M | -0.76 | M | -0.52 | M |
| Giresun | -1.23 | M | -1.24 | M | -1.14 | M | 0.20 | M | -0.52 | M | 0.75 | M | -0.39 | M | -0.52 | M | -0.19 | M |
| Gümüşhane | -1.50 | L | -1.36 | L | -1.33 | L | -0.21 | M | -0.53 | M | -0.80 | M | -0.65 | M | -0.53 | M | -0.89 | M |
| Hakkari | -2.32 | L | -2.46 | L | -2.29 | L | -3.69 | L | -1.92 | L | -2.14 | L | -2.31 | L | -1.92 | L | -2.14 | L |
| Hatay | 0.59 | M | 0.37 | M | 0.12 | M | -1.25 | M | -0.38 | M | -0.94 | M | -0.26 | M | -0.38 | M | -0.32 | M |
| Iğdır | -2.12 | L | -2.10 | L | -1.88 | L | -2.25 | L | -1.80 | L | -2.23 | L | -1.68 | L | -1.80 | L | -1.70 | L |
| Isparta | 0.32 | M | 0.19 | M | 0.29 | M | 3.62 | VH | 1.02 | H | 2.93 | VH | 1.52 | H | 1.02 | H | 1.30 | H |
| İstanbul | 7.99 | VH | 7.77 | VH | 8.19 | VH | 2.84 | VH | 4.03 | VH | 2.76 | VH | 4.15 | VH | 4.03 | VH | 4.61 | VH |
| İzmir | 4.67 | VH | 3.95 | VH | 2.99 | VH | 3.26 | VH | 2.70 | VH | 3.07 | VH | 3.35 | VH | 2.70 | VH | 2.51 | VH |
| Kahramanmaraş | -0.23 | M | -0.25 | M | -0.06 | M | -1.50 | L | -0.70 | M | -1.49 | L | -0.66 | M | -0.70 | M | -0.62 | M |
| Karabük | 0.02 | M | 0.24 | M | 0.52 | M | 2.63 | VH | 0.49 | M | 2.25 | H | 1.02 | H | 0.49 | M | 1.12 | H |
| Karaman | 0.35 | M | -0.05 | M | 0.31 | M | -1.23 | M | -0.38 | M | -0.55 | M | -0.34 | M | -0.38 | M | -0.09 | M |
| Kars | -1.98 | L | -1.81 | L | -1.84 | L | -1.60 | L | -1.33 | L | -1.59 | L | -1.37 | L | -1.33 | L | -1.18 | M |
| Kastamonu | -0.19 | M | 0.10 | M | 0.08 | M | 0.28 | M | -0.11 | M | 0.46 | M | 0.03 | M | -0.11 | M | 0.22 | M |
| Kayseri | 0.92 | M | 0.89 | M | 0.60 | M | 0.86 | M | 0.96 | M | 0.56 | M | 0.68 | M | 0.96 | M | 0.48 | M |
| Kırıkkale | -1.02 | M | -0.43 | M | -0.40 | M | 0.99 | M | 0.67 | M | 1.97 | H | -0.01 | M | 0.67 | M | 0.62 | M |
| Kırklareli | 1.23 | H | 1.29 | H | 1.47 | H | 1.91 | H | 1.23 | H | 2.12 | H | 1.21 | H | 1.23 | H | 1.48 | H |
| Kırşehir | -0.73 | M | -0.35 | M | -0.34 | M | 0.31 | M | -0.11 | M | -0.01 | M | -0.16 | M | -0.11 | M | -0.15 | M |
| Kilis | -1.76 | L | -1.63 | L | -1.17 | M | -1.25 | M | -1.38 | L | -0.80 | M | -1.16 | M | -1.38 | L | -0.82 | M |
| Kocaeli | 2.80 | VH | 2.98 | VH | 2.54 | VH | 0.97 | M | 1.57 | H | 1.37 | H | 1.44 | H | 1.57 | H | 1.64 | H |
| Konya | 2.46 | H | 1.44 | H | 2.09 | H | -0.38 | M | 0.72 | M | 0.43 | M | 0.79 | M | 0.72 | M | 1.07 | H |
| Kütahya | -0.05 | M | 0.19 | M | 0.19 | M | 0.37 | M | 0.12 | M | 1.36 | H | 0.12 | M | 0.12 | M | 0.63 | M |
| Malatya | -0.53 | M | -0.86 | M | -0.75 | M | 0.15 | M | 0.06 | M | 0.89 | M | -0.15 | M | 0.06 | M | 0.08 | M |
| Manisa | 1.53 | H | 1.39 | H | 1.17 | H | 0.10 | M | 0.59 | M | 0.46 | M | 0.62 | M | 0.59 | M | 0.68 | M |
| Mardin | -2.00 | L | -1.69 | L | -1.44 | L | -3.95 | L | -1.92 | L | -3.68 | L | -2.29 | L | -1.92 | L | -2.09 | L |
| Mersin | 1.48 | H | 1.15 | H | 1.41 | H | 0.14 | M | 0.36 | M | 0.22 | M | 0.62 | M | 0.36 | M | 0.69 | M |
| Muğla | 2.01 | H | 1.84 | H | 1.23 | H | 1.47 | H | 0.87 | M | 1.24 | H | 1.34 | H | 0.87 | M | 1.02 | H |
| Muş | -2.09 | L | -2.06 | L | -2.14 | L | -4.16 | L | -2.25 | L | -3.23 | L | -2.40 | L | -2.25 | L | -2.21 | L |
| Nevşehir | -0.07 | M | -0.24 | M | -0.28 | M | 0.06 | M | -0.20 | M | -1.13 | M | 0.00 | M | -0.20 | M | -0.58 | M |
| Niğde | -0.33 | M | -0.51 | M | -0.14 | M | -0.80 | M | -0.73 | M | -1.42 | L | -0.43 | M | -0.73 | M | -0.63 | M |
| Ordu | -0.77 | M | -1.15 | M | -0.85 | M | -0.44 | M | -0.78 | M | 0.02 | M | -0.46 | M | -0.78 | M | -0.36 | M |
| Osmaniye | -1.06 | M | -0.40 | M | 0.41 | M | -1.03 | M | -0.19 | M | -0.90 | M | -0.38 | M | -0.59 | M | -0.36 | M |
| Rize | -0.78 | M | -0.65 | M | -0.62 | M | 0.42 | M | -0.05 | M | 0.39 | M | -0.13 | M | -0.05 | M | -0.11 | M |
| Sakarya | 1.12 | H | 0.87 | M | 0.48 | M | -0.40 | M | 0.06 | M | 0.05 | M | 0.27 | M | 0.06 | M | 0.22 | M |
| Samsun | 0.60 | M | 0.19 | M | 0.07 | M | 0.58 | M | 0.43 | M | 0.45 | M | 0.45 | M | 0.43 | M | 0.30 | M |
| Siirt | -2.09 | L | -1.96 | L | -1.93 | L | -2.81 | L | -1.71 | L | -2.48 | L | -1.89 | L | -1.71 | L | -1.82 | L |
| Sinop | -0.78 | M | -0.54 | M | -0.93 | M | 0.99 | M | -0.46 | M | -1.31 | L | 0.06 | M | -0.46 | M | -0.92 | M |
| Sivas | -0.36 | M | -0.33 | M | -0.38 | M | 0.39 | M | 0.35 | M | 0.68 | M | 0.00 | M | 0.35 | M | 0.11 | M |
| Şanlıurfa | -0.72 | M | -0.67 | M | -0.67 | M | -2.44 | L | -1.57 | L | -4.68 | L | -1.22 | M | -1.57 | L | -2.16 | L |
| Şırnak | -2.35 | L | -2.48 | L | -2.22 | L | -4.26 | L | -2.23 | L | -3.66 | L | -2.54 | L | -2.23 | L | -2.42 | L |
| Tekirdağ | 2.68 | VH | 2.51 | VH | 2.22 | H | 1.00 | H | 1.31 | H | 0.98 | M | 1.41 | H | 1.31 | H | 1.34 | H |
| Tokat | -0.86 | M | -0.83 | M | -0.79 | M | -0.10 | M | -0.43 | M | 0.04 | M | -0.37 | M | -0.41 | M | -0.35 | M |
| Trabzon | -0.44 | M | -0.57 | M | -0.57 | M | 2.26 | H | 0.62 | M | 1.31 | H | 0.70 | M | 0.62 | M | 0.28 | M |
| Tunceli | -1.51 | L | -1.29 | M | -1.08 | M | 0.78 | M | 0.05 | M | 1.06 | H | -0.27 | M | 0.05 | M | -0.04 | M |
| Uşak | 0.04 | M | 0.27 | M | 0.54 | M | 0.47 | M | 0.20 | M | 0.77 | M | 0.22 | M | 0.20 | M | 0.30 | M |
| Van | -1.91 | L | -2.00 | L | -2.28 | L | -2.42 | L | -1.79 | L | -2.95 | L | -1.66 | L | -1.79 | L | -2.16 | L |
| Yalova | 0.46 | M | 0.37 | M | 0.63 | M | 0.81 | M | 0.84 | M | 0.62 | M | 0.49 | M | 0.84 | M | 0.68 | M |
| Yozgat | -0.97 | M | -0.89 | M | -0.92 | M | -0.87 | M | -0.73 | M | -0.97 | M | -0.71 | M | -0.73 | M | -0.78 | M |
| Zonguldak | -0.03 | M | 0.18 | M | 0.26 | M | 1.29 | H | 0.97 | M | 1.74 | H | 0.49 | M | 0.97 | M | 0.81 | M |

Low    Medium    High    Very High 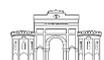

**Figure 9**
*Socioeconomic Disparities Map of Türkiye*

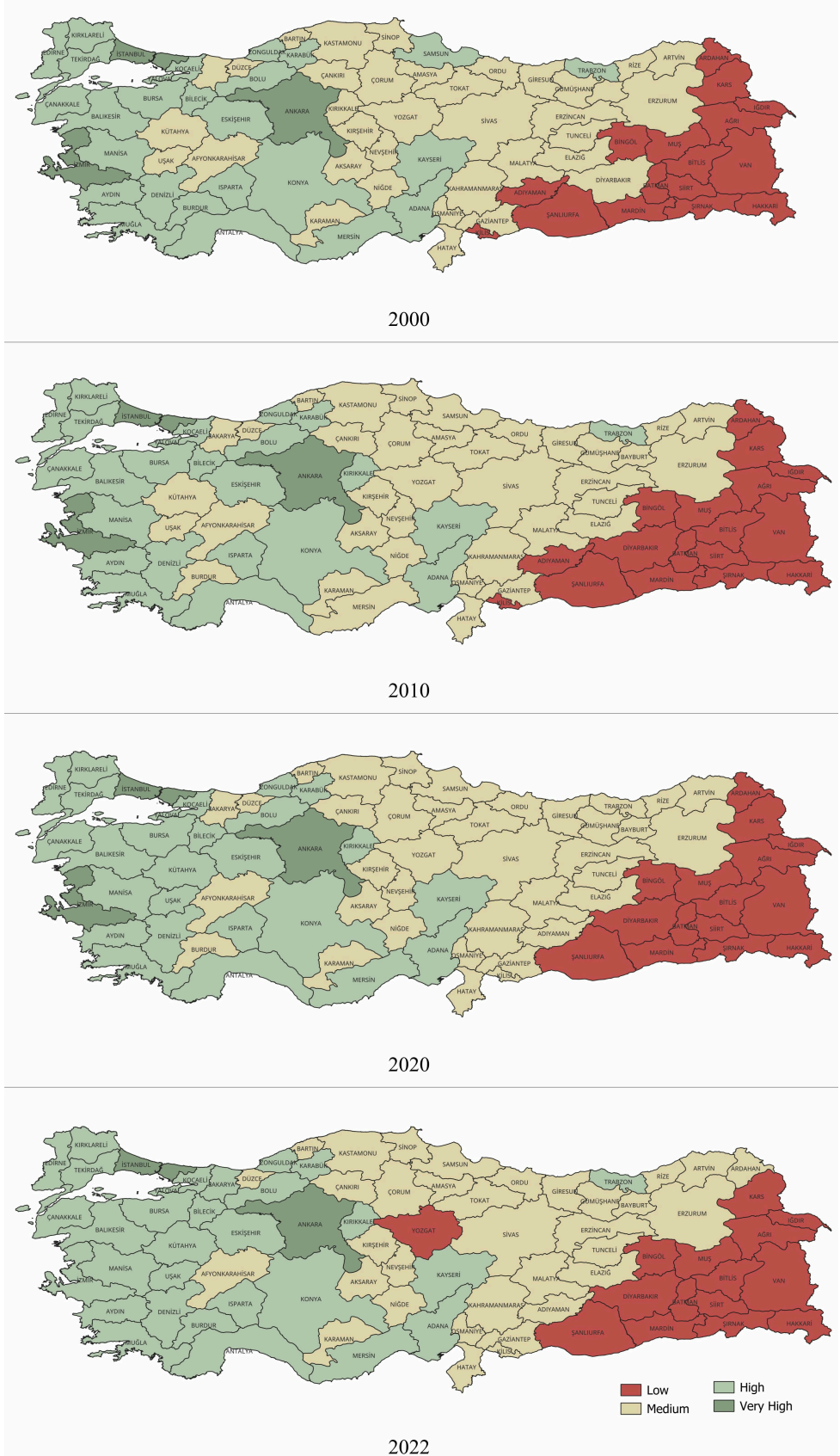

2000

2010

2020

2022

Low    High
Medium    Very High

Figure 8 shows the social, economic and socioeconomic indices of the provinces in Türkiye in 2000, 2010, and 2020. Maps of these indices are also shown in Figure 9. The data shows that there have been significant changes in the economic and social areas. These changes can be explained by various factors such as the internal dynamics of provinces, local government policies and national economic conditions.

Based on the clustering results, the economic index emphasizes provinces such as Burdur and Edirne, which moved from the medium cluster to the high cluster. These provinces showed a significant economic increase from 2000 to 2010 but declined to the medium cluster by 2020. Meanwhile, provinces in the high cluster, such as Adana, Aydın, Bolu, Kayseri and Sakarya, have fallen to the medium cluster by 2020. Such declines may indicate economic difficulties and perhaps a reduction in investment. Moreover, provinces such as Diyarbakır and Ordu dropped from the medium cluster to the low cluster, indicating even greater economic challenges. Interestingly, a large city like Izmir fell from the very high cluster to the high cluster, indicating that metropolitan provinces can also be affected by economic fluctuations. Similar changes were observed in the social index. Provinces such as Afyonkarahisar, Bayburt, Ordu and Sakarya moved from the medium to the high cluster, indicating improvements in their social infrastructure and quality of life. Provinces such as Bolu, Çanakkale, Kırıkkale, Kırklareli and Zonguldak moved from high to very high, reaching the top cluster. However, many provinces dropped from the high cluster to the medium cluster, which may indicate weakening of social services and infrastructure. In particular, provinces such as Edirne, Eskişehir, Isparta and Karabük dropped from the very high cluster to the high cluster, indicating that even the best provinces can face social challenges. The socioeconomic index analysis also revealed important findings. Kütahya and Uşak moved up from the medium to the high cluster by combining their economic and social development. However, provinces such as Burdur, Samsun and Mersin have fallen from the high cluster to the medium cluster. These declines may be an indicator of socioeconomic instability. Provinces such as Adıyaman and Kilis have made some progress, moving from low to medium clusters. Large cities such as Ankara, Istanbul and Izmir have consistently remained in the very high cluster, maintaining their position as the most developed and stable regions in Türkiye. However, Izmir was found to fall from the very high cluster to the high cluster in 2022.

The clustering results clearly reveal the geographical distribution of the socioeconomic inequalities in Türkiye. Provinces in the lower cluster such as Ağrı, Hakkari and Siirt are generally less economically developed and have lower values in terms of social indicators. Living standards and income levels in these provinces are generally below the Türkiye average. In particular, the provinces in Eastern and Southeastern Anatolia face serious problems in terms of socioeconomic indicators. Provinces such as Afyonkarahisar, Amasya and Artvin are in the medium cluster. In these provinces, economic activities are mainly based on agriculture, livestock, and small-scale industry. The accessibility and quality of education and health services as well as the state of infrastructure have also contributed to their placement in the medium cluster. These provinces reflect the average socioeconomic performance of Türkiye. These results are similar to Çakmak and Örkçü (2016), and Arı and Hüyüktepe (2019).

Provinces with high socioeconomic indices exhibit superior performance in terms of economic and social indicators. Provinces such as Adana, Antalya and Aydın fall into this cluster. In these provinces, tourism and agriculture-based economic activities increase per capita income and living standards. In particular, Antalya and Aydın have high socioeconomic indicators due to tourism. These provinces have well-developed infrastructure and health services. The provinces in the very high cluster represent the most developed provinces in Türkiye. These provinces have a competitive performance in terms of socioeconomic indicators not only at the national but also at the international level. Ankara and Istanbul stand out as the two most socioeconomically developed provinces in Türkiye. They have a high GDP per capita, wide access to health and education services, and well-developed infrastructure. This shows that these cities are the locomotive of the country in terms of both economic activity and quality of life. These results are consistent with Özkubat and Selim (2019), Servi and Erişoğlu (2020), and Karadaş and Erilli (2023).

There is a consistency between the results obtained from the k-means clustering analysis and the results of the clusters provided by the dendrogram. These results are particularly consistent with clusters based on the socioeconomic structure of cities. It is noteworthy that the socioeconomic factors that are evident in the K-means analysis tend to cluster cities similarly in the clustering using the dendrogram. This suggests that both methods are valid and robust for clustering the socioeconomic structure of the provinces in Türkiye. The results of the analysis provide valuable information for shaping regional policies and identifying development strategies. In conclusion, the clustering of provinces in Türkiye based on socioeconomic indicators reveals that the socioeconomic structure of the country is complex and diverse. The results of the study will contribute to the development of policies for the socioeconomic development of Türkiye.

5. Conclusions

Economic development and regional disparities are important issues for comprehending and managing the socioeconomic structure of a country. In a geographically, culturally and economically diverse country like Türkiye, determining the socioeconomic status of the provinces and regional disparities is an important step for effective policy planning and implementation. Therefore, in this study, a cluster analysis was conducted to determine the socioeconomic disparities of the provinces in Türkiye. The results of the analysis provided a detailed examination of the country's regional differences and levels of economic development. The results can provide an important reference for the formulation and implementation of economic and social policies in Türkiye.

An economic analysis of the provinces in Türkiye reveals that large metropolitan provinces such as Istanbul, Ankara and Izmir are the economic engines of Türkiye. These provinces are characterized by high economic index values, large industrial bases, high income levels, and intensive trade activities. Istanbul in particular has a much higher economic index compared to other provinces. In contrast, the provinces in Eastern Anatolia and Southeastern Anatolia are less economically developed. This should be considered when guiding regional economic policies. In terms of social index, provinces such as Ankara, Istanbul and Edirne have high social indexes. It is seen that living standards and social services are developed in these provinces. However, provinces with a low social index are less advantageous in terms of access to education and health services, employment opportunities and overall quality of life. Special efforts are needed to improve social welfare in these provinces.

Comparative analysis of the socioeconomic index shows that provinces such as Kütahya and Uşak have moved from the medium cluster to the high cluster in the 2000-2020 period. This transition shows that significant progress has been made in both economic and social areas and that these provinces have developed holistically. In contrast, provinces such as Burdur, Samsun and Mersin declined from high to medium cluster. These declines indicate that the socioeconomic conditions in these provinces have become more difficult and that these regions are experiencing some difficulties in terms of social infrastructure as well as economic potential. In particular, factors such as economic recessions or disruptions in social services can be considered among the main reasons for this decline. Provinces such as Adıyaman and Kilis have risen from the low cluster to the medium cluster. These provinces have made significant progress in terms of their economic situation and access to social services. Such an increase reflects the positive results of the policies implemented by the local authorities and the central government. In addition, Türkiye's metropolitan cities of Ankara, Istanbul and Izmir have consistently remained in the very high cluster. This indicates that these metropolitan provinces have maintained their position as the most developed regions in Türkiye, preserving both their economic strength and social infrastructure. These provinces have been able to maintain a high level of socioeconomic stability and continuous development. However, Izmir declined from the Very High cluster to the High cluster in 2022.

The 2022 socioeconomic cluster analysis results show that only two provinces in Türkiye have a very high socioeconomic status. This reveals that the number of provinces with very high socioeconomic status across the country is extremely limited. Ankara and Istanbul, which are in the very high cluster, perform competitively at both national and international levels with high GDP per capita, advanced health and education services, and widely accessible infrastructure. These provinces play the role of the country's locomotive in terms of economic activity and quality of life. However, the 30 provinces in the high socioeconomic cluster account for 37% of all the provinces in Türkiye. This shows that many regions of the country can reach high socioeconomic levels, but the number of these provinces-reaching the highest level is limited. Provinces in the high cluster are regions that perform well in terms of economic and social indicators, where tourism and agriculture-based economic activities have raised per capita income and living standards. The development of infrastructure and health services in these provinces generally supports high socioeconomic development. Provinces in the medium and low socioeconomic cluster account for almost one-third of Türkiye, indicating the existence of socioeconomic inequalities and imbalances in the country. The provinces in the medium cluster are those where economic activities are mostly based on agriculture, livestock, and small-scale industry. Provinces in the low cluster are generally less economically developed and have lower values in terms of social indicators, with living standards and income levels generally below the national average.

This analysis concludes that Türkiye's development policies should be designed considering the socioeconomic status of the provinces and regional disparities. For this purpose, policies have been developed for Türkiye's regional development and the reduction of socioeconomic disparities. These policy recommendations are designed to reduce Türkiye's regional disparities, promote economic and social development and increase the overall welfare of the country. The policy recommendations can be listed as follows: (1) To comprehend the economic and social needs in various regions of Türkiye and to develop policies in line with these needs, it is recommended to increase the capacity and restructure the activities of regional development agencies. These agencies can implement region-specific projects by assessing local potential and collaborating with local actors. Moreover, these agencies should play an active role in experience sharing and knowledge transfer to disseminate successful policies to other provinces. (2) Improving access to education and health services should be a priority target in regions with low social index values. Increasing the quality and accessibility of education and expanding and improving health services will contribute to improving the quality of life and social welfare in these regions. (3) Infrastructure investments should be accelerated in regions lagging behind in terms of economic and social development. Improvement of basic infrastructure services such as transportation, energy, communication and water can revitalize economic activities and improve the social quality of life in these regions. Infrastructure investments in these provinces should be supported by crisis management units that can respond swiftly in times of crisis. (4) To reduce regional economic disparities, it is important to diversify local economies and support small and medium-sized enterprises (SMEs). This should be especially in sectors such as agriculture, tourism and renewable energy that can mobilize local potential. Regional development projects and local economic activities should be supported in provinces with development potential such as Adıyaman and Kilis. (5) Innovation and technology investments should be encouraged to boost economic development and competitiveness. In particular, supporting high-tech ecosystems, research and development centers and start-up ecosystems will accelerate regional development. (6) Investments in technology and innovation should be increased to ensure that metropolitan cities such as Ankara, Istanbul and Izmir remain at consistently high socioeconomic levels. These cities should be planned sustainably with population planning and increased infrastructure. Innovative solutions to metropolitan problems such as traffic, environmental pollution and housing prob-lems should be developed. (7) Programs to increase social integration and employment should be developed in regions with low social index values. These programs should provide education, skills development and employment opportunities for youth, women and disadvantaged groups. (8) Prevent migration movements that would disrupt social integration. External migration should be limited and planned. Social and economic integration programs should be developed to ensure the integration of migrants. These programs should be designed to support the integration of migrants into local communities. (9) The capacity of regional and local governments to plan and implement economic and social development projects in their regions should be enhanced. This will allow for more effective and localized policy-making processes. (10) International cooperation and investment should be encouraged to support Türkiye's economic and social development. This can be achieved through projects and funds, particularly from the European Union and other international organizations. (11) Central government policies should be revised to ensure equitable distribution of resources among provinces, data-based policies should be developed and the participation of local governments in decision-making processes should be increased.

This study has important implications. However, this study also has some limitations. Future studies can improve addressing these limitations. First, this study includes 16 indicators under the economic and social dimensions. The study can be improved by increasing the dimensions and indicators. Second, Türkiye and 81 provinces were selected as the sample in this study. The sample of future studies can be selected as developed and developing countries, and the results obtained can be compared. Third, the elbow and k-means methods were used in this study. The results obtained in this study can be compared using different and/or hybrid techniques.

| | |
|---|---|
| Peer Review | Externally peer-reviewed. |
| Conflict of Interest | The author has no conflict of interest to declare. |
| Grant Support | The author declared that this study has received no financial support. |
| Author Details | **Emre Akusta (Assist. Prof.)** |
| | ¹ Kırklareli University, Faculty of Economics and Administrative Sciences, Department of Economics, Kırklareli, Türkiye |
| |  0000-0002-6147-5443   ✉ emre.akust@klu.edu.tr |